\documentclass
[twocolumn,aps,prc,amsmath,showpacs,amssymb,floatfix]
{revtex4}
\usepackage{CJK}                     
\usepackage[dvips]{graphicx}
\usepackage{bm}                      
\usepackage{mathptmx}                
\usepackage{dcolumn}                 

\begin{document}

\def\bea{\begin{eqnarray}}
\def\eea{\end{eqnarray}}
\def\be{\begin{equation}}
\def\ee{\end{equation}}
\def\rra{\right\rangle}
\def\lla{\left\langle}
\def\rv{\bm{r}}
\def\eps{\epsilon}
\def\intdp{\int\!\!\frac{d^3p}{(2\pi)^3}}


\title{
Protoneutron stars
in the Brueckner-Hartree-Fock approach\\
and finite-temperature kaon condensation}

\begin{CJK}{GB}{gbsn}


\author{A. Li (Àî°º) and X. R. Zhou (ÖÜÏÈÈÙ)}

\affiliation{
Institute of Theoretical Physics and Astrophysics, Department of Physics, 
Xiamen University, Xiamen 361005, P.~R.~China}

\author{G. F. Burgio and H.-J. Schulze}
\affiliation{
INFN, Sezione di Catania, Via Santa Sofia 64, I-95123 Catania, Italy}


\begin{abstract}
We study the properties of hot neutrino-trapped $\beta$-stable stellar matter
using an equation of state of nuclear matter
within the Brueckner-Hartree-Fock approach including three-body forces,
combined with a standard chiral model for kaon condensation at finite temperature. 
The properties of (proto)neutron stars are then investigated
within this framework.
\end{abstract}


\maketitle
\end{CJK}

\section{Introduction}

One of the challenging problems in nuclear physics is to elucidate the behavior
of nuclear matter in high-density and/or high-temperature environments,
particularly relevant for compact stellar objects
like (proto)neutron stars [(P)NS].
Despite the importance of the structure and properties of $\beta$-stable matter
at extreme densities of several times normal nuclear matter density
($\rho_0 \approx 0.17\;\text{fm}^{-3}$),
its internal constitution and the equation of state (EOS) are not
yet known with certainty.

At such densities strangeness may occur in the form of hadrons
(such as hyperons or a $K^-$ meson condensate)
or in the form of strange quarks.
The existence of these strange matter phases may have important
consequences for the structure of compact stars and for the cooling
dynamics of the PNS after a supernova explosion.
With respect to kaons, the suggestion of Kaplan and Nelson \cite{kaplan}
that at high enough densities the ground state of baryonic matter might contain
a Bose-Einstein (BE) condensate of negatively charged kaons has
prompted extensive investigations and discussions
\cite{politzer,brownevo,brown,thorsson,ellis,glendenning,
rep,tatsumi,tatsumievo,pons,ramos,carlson,norsen,kubis,liang1,liang2}
on its implications for astrophysical phenomena in (P)NS's.
In particular, the proton abundance is increased dramatically when a
kaon condensate is present in NS matter,
and antileptons are allowed to exist.

Some authors treated kaon condensation within an improved chiral perturbation
theory beyond the tree-order calculations, 
and their results indicated that the
critical density $\rho^K_c$ for kaon condensation lies in the range of
$2\rho_0 \lesssim \rho^K_c \lesssim 4\rho_0$.
The critical density depends sensitively on the value of the strangeness
content of the proton, which is still quite
controversial \cite{thorsson,rep,kubis,oldlat,dong,gus,lyubovitskij,ohki}.

Estimates of the relevant formation timescales \cite{brownevo,tatsumievo}
indicate that the build-up of the kaon condensate is very fast
compared to the typical cooling and neutrino-diffusion timescale
of several seconds characteristic for a PNS,
and could even play a role during the preceding supernova core collapse.
Therefore, a kaon condensate might be present immediately after the
formation of a PNS, and influence its evolutionary history.
In fact, lepton trapping in a PNS shifts the onset of kaon
condensation to higher densities as compared to neutrino-free matter.
Also the presence of other strange particles (for instance, hyperons)
was found to push the onset of kaon condensation to higher densities,
even out of the physically relevant density regime, 
$\rho\lesssim 1\;\text{fm}^{-3}$, \cite{rep,schaffner}.
This leads to the widely discussed possibility of a delayed collapse
of the PNS to a black hole,
when during the cooling and deleptonization evolution
the increasing softening effect of the kaons on the EOS becomes too big
to stabilize an initially very massive star \cite{brown,rep,tatsumi,pons,baum}.

Obviously, for a reliable modeling of this effect
both conditions of finite temperature and lepton trapping have to
be taken into account.
However, most of the previous investigations have been done for cold matter,
thus neglecting the dependence of the kaon condensation on temperature,
which plays a role in affecting significantly the properties of
PNS's \cite{taka,schaab,nbbs,isen}.
Therefore we extend our previous work \cite{liang1,liang2} to hot matter.
The main goal of this article is to investigate the impact of a kaon condensate
on PNS matter at finite temperature and on the final PNS observables,
combining a microscopic Brueckner-Hartree-Fock (BHF) approach
for the baryonic part of the matter
with a standard chiral model for the kaon-nucleon contribution.

Our paper is organized as follows.
In Sec.~IIA we discuss the finite-temperature BHF approach,
and in Sec.~IIB the standard chiral model at finite temperature.
The composition of stellar matter and the EOS are presented in Sec.~III,
along with the equations of stellar structure.
The numerical results are then illustrated in Sec.~IV,
and conclusions are drawn in Sec.~V.

\section{Theoretical Models}

In the kaon-condensed phase of (P)NS matter,
the free energy density consists of three contributions,
\be
 f = f_{NN} + f_{KN} + f_{L} \:,
\label{e:f}
\ee
where $f_{NN}$ is the baryonic part,
$f_{KN}$ is the kaonic part including the contribution
from the kaon-nucleon interaction, and
$f_{L}$ denotes the contribution of leptons $e,\mu,\nu_e,\nu_\mu$,
and their antiparticles.

\subsection{Brueckner-Bethe-Goldstone theory at finite temperature}

\begin{figure}
\includegraphics[height=70mm,angle=0]{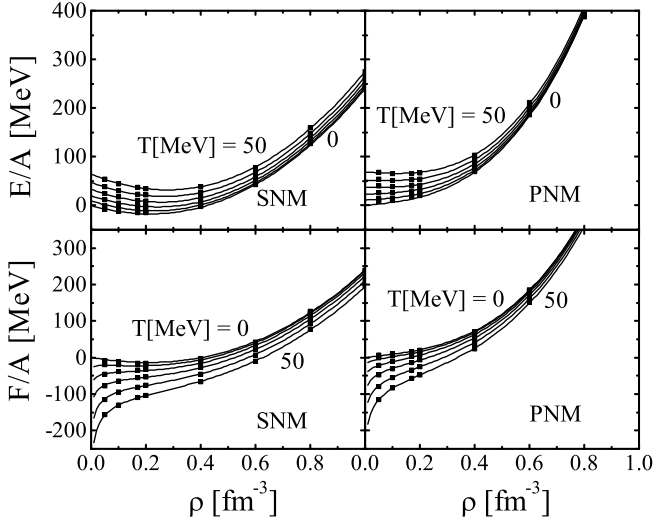}
\caption{
Finite-temperature EOS for symmetric (left panels)
and purely neutron (right panels) matter.
The internal energy (upper panels) and the free energy
(lower panels) are displayed as a function of the nucleon density,
for temperatures ranging from 0 to 50 MeV in steps of 10 MeV.
The numerical data (markers) and the results of the fits,
Eqs.~(\ref{e:fite},\ref{e:fitf}), (curves) are shown.}
\label{f:bf}
\end{figure}

In the present work, we employ the BHF approach
for asymmetric nuclear matter at finite
temperature \cite{nbbs,lej,bombaci,book,baldo} to
calculate the baryonic contribution to the EOS of stellar matter.
The essential ingredient of this approach is the interaction matrix $G$,
which satisfies the self-consistent equations
\be
 G(\rho,x;E) = V + V \sum_{1,2}
 \frac{|12 \rangle (1-n_1)(1-n_2) \langle 1 2|}
 {E - e_1-e_2 +i0} G(\rho,x;E) \:
\label{eq:BG}
\ee
and
\be
 U_1(\rho,x) = {\rm Re} \sum_2 n_2
 \langle 1 2| G(\rho,x;e_1+e_2) | 1 2 \rangle_a \:,
\label{eq:uk}
\ee
where $x=\rho_p/\rho$ is the proton fraction, and
$\rho_p$ and $\rho$ are the proton and the total baryon density, respectively.
$E$ is the starting energy and
$e(k) \equiv k^2\!/2m + U(k)$ is the single-particle (s.p.) energy.
The multi-indices 1,2 denote in general momentum, isospin, and spin.

The realistic nucleon-nucleon ($NN$) interaction $V$
adopted in the present calculation is the
Argonne $V_{18}$ two-body force \cite{wiringa}
supplemented by either a microscopic three-body force (TBF)
based on the meson-exchange approach \cite{grange,zuotbf,litbf}
(denoted micro TBF in the following),
or the phenomenological Urbana UIX force discussed in Refs.~\cite{uix,bbb}
(pheno TBF),
which are reduced to an effective two-body force and added to the bare potential
in the BHF calculation
(see Refs.~\cite{grange,zuotbf,litbf} for details).

At finite temperature, $n(k)$ in Eqs.~(\ref{eq:BG}) and (\ref{eq:uk})
is a Fermi distribution.
For a given density and temperature, these equations have
to be solved self-consistently along with the following equations for
the auxiliary chemical potentials $\tilde{\mu}_{n,p}$ ,
\be
 \rho_i = 2\sum_k n_i(k) =
 2\sum_k {\left[\exp{\Big(\frac{e_i(k)-\tilde{\mu}_i}{T}\Big)}
 + 1 \right]}^{-1} \:.
\label{eq:ro}
\ee

To save computational time and simplify the numerical procedure,
in the following we employ the so-called
{\it Frozen Correlations Approximation} \cite{nbbs,baldo}, i.e.,
the correlations at $T \neq 0$
are assumed to be essentially the same as at $T = 0$.
This means that the s.p.~potential $U_i(k)$ for the component $i$
at finite temperature is approximated by the one calculated at $T=0$.
Within this approximation,
the nucleonic free energy density has the following simplified expression,
\be
 f_{NN} = \sum_{i=n,p} \left[ 2\sum_k n_i(k)
 \left( {k^2\over 2m_i} + {1\over 2}U_i(k) \right) - Ts_i \right] \:,
\ee
where
\be
 s_i = - 2\sum_k \Big( n_i(k) \ln n_i(k) + [1-n_i(k)] \ln [1-n_i(k)] \Big)
\ee
is the entropy density for the component $i$ treated as a free Fermi gas with
spectrum $e_i(k)$.
It turns out that the assumed
independence is valid to a good accuracy \cite{nbbs,baldo},
at least for not too high temperature, $T\lesssim30\;\text{MeV}$.

For illustration, we display in Fig.~\ref{f:bf} the EOS obtained following
the above discussed procedure, for symmetric nuclear matter and purely
neutron matter, adopting the micro TBF.
In the upper panels we display the internal energy per particle,
whereas the lower panels show the free energy as a function of the
baryon density, for several values of temperature between 0 and 50 MeV.
We notice that the free energy of symmetric matter is a monotonically
decreasing function of temperature.
At $T=0$ the free energy coincides with the internal energy
and the corresponding curve is just the usual nuclear matter saturation curve.
On the contrary, the internal energy is an increasing function of temperature.
The effect is less pronounced for pure neutron matter due to the larger
Fermi energy of the neutrons at given density.
We notice that the results of this microscopic TBF are always stiffer
than those of the phenomenological Urbana TBF \cite{uix}
used in Ref.~\cite{bbb}.

For practical use, we provide analytical fits of the internal energy
$E/A(T,\rho,x)$ as well as the free energy $F/A(T,\rho,x)$.
It turns out that for both quantities the dependence on proton fraction
can be very well approximated by a quadratic dependence,
as at zero temperature \cite{bombaci,bbs}:
\be
 {E\over A}(T,\rho,x) \approx
 {E\over A}(T,\rho,x=0.5) + (1-2x)^2 E_\mathrm{sym}(T,\rho) \:,
\label{e:parab}
\ee
where the symmetry energy $E_\mathrm{sym}$ can be expressed in
terms of the difference of the energy per particle between pure neutron
($x=0$) and symmetric ($x=0.5$) matter:
\bea
 E_\mathrm{sym}(T,\rho) &=&
 - {1\over 4} {\partial(E/A) \over \partial x}(T,\rho,0)
\\
 &\approx& {E\over A}(T,\rho,0) - {E\over A}(T,\rho,0.5) \:.
\label{e:sym}
\eea
Therefore, it is only necessary to provide parametrizations
of both quantities for symmetric nuclear matter and pure neutron matter.
We find that the following functional forms provide excellent parametrizations
of the numerical results in the required ranges of density 
($0.03\;\text{fm}^{-3} \lesssim \rho \lesssim 1\;\text{fm}^{-3}$)
and temperature ($0\;\text{MeV} \leq T \leq 50\;\text{MeV}$):
\bea
 {E\over A}(\rho,T)&=& (a_1t+a_2t^2) + (b_0+b_1t)\rho + (c_0+c_1t)\rho^d \:,\quad
\label{e:fite}
\\
 {F\over A}(\rho,T)&=& (a_1t+a_2t^2)\ln(\rho) + (b_0+b_2t^2)\rho + c_0\rho^d \:,
\quad
\label{e:fitf}
\eea
where $t=T/(100\;\text{MeV})$ and $E,F$, and $\rho$ are given in
MeV and fm$^{-3}$, respectively.
The parameters of the different fits are given in Table~\ref{t:fit}
for both TBF's we are using.

\begin{table}
\caption{
 Parameters of the EOS fits, Eqs.~(\ref{e:fite},\ref{e:fitf}),
 for symmetric nuclear matter (SNM) and pure neutron matter (PNM)
 and both nuclear TBF's used.}
\medskip
\begin{ruledtabular}

\begin{tabular}{l|rrrccccc}
 micro TBF   &$a_1$ &$a_2$ &$b_0$ &$b_1$ &$b_2$ &$c_0$ &$c_1$ &$d$ \\
\hline
 $\!\!\!\!E/A$, SNM &  81  & 95   & -155 & -139 &      & 395  & 81   & 2.09 \\
 $\!\!\!\!E/A$, PNM & 101  & 73   &   54 & -181 &      & 659  & 84   & 2.88 \\
 $\!\!\!\!F/A$, SNM & 41   & 120  & -115 &      & -182 & 355  &      & 2.24 \\
 $\!\!\!\!F/A$, PNM & 18   & 123  &  83  &      & -103 & 631  &      & 3.02 \\
\end{tabular}

\begin{tabular}{l|rrrccccc}
 pheno TBF   &$a_1$ &$a_2$ &$b_0$ &$b_1$ &$b_2$ &$c_0$ &$c_1$ &$d$ \\
\hline
 $\!\!\!\!E/A$, SNM & 105  & 74   & -473 & -464 &      & 586  & 381  & 1.26 \\
 $\!\!\!\!E/A$, PNM & 109  & 64   &   34 & -240 &      & 249  & 164  & 1.97 \\
 $\!\!\!\!F/A$, SNM & 41   & 116  & -180 &      & -174 & 293  &      & 1.57 \\
 $\!\!\!\!F/A$, PNM & 21   & 116  &  101 &      & -131 & 191  &      & 2.62 \\
\end{tabular}

\end{ruledtabular}
\label{t:fit}
\end{table}

\subsection{Kaon condensate at finite temperature}

Kaon condensation in nuclear matter has been studied intensively in a large
variety of models.
For the required extension to finite temperature we employ the formalism of
Refs.~\cite{tatsumi,tatsumievo}, which treats
fluctuations around the condensate within the framework of chiral symmetry.
For small condensate amplitudes this approach is exactly equivalent to the
meson-exchange mean-field models of Ref.~\cite{pons},
and we briefly review it now.

In the following equations, $m_K$ and $\mu_K$ are the kaon mass
and chemical potential,
$f_\pi=93\;\text{MeV}$ is the pion decay constant,
$\theta$ is the amplitude of the condensate,
\be
 E^\pm_p =
 \sqrt{p^2 + {\widetilde m}_K^2} \pm \widetilde\mu_K
\label{e:epm}
\ee
are the kaonic excitation energies \cite{thorsson,tatsumi,pons,kubis}
with
\bea
 \widetilde m_K &=& \sqrt{{m_K^*}^2\cos\theta + b^2} \:,
\label{e:mk}
\\
 \widetilde\mu_K &=& \mu_K\cos\theta + b \:,
\label{e:muk}
\eea
and
\bea
 {m_K^*}^2 &=& m_K^2 + (a_1x+a_2+2a_3)m_s\rho/f_\pi^2 \:,
\label{e:mkstar}
\\
 b &=& (1+x)\rho/4f_\pi^2
\eea
are the scalar effective kaon mass and the $V$-spin density, respectively.

We adopt the `standard' $KN$ interaction parameters
\cite{politzer,thorsson,rep,tatsumi,kubis}
$a_1m_s=-67$ MeV, $a_2m_s=134$ MeV, and
$a_3m_s =-134$, $-222$, $-310$ MeV
to perform our numerical calculations,
where the different choices of $a_3$ correspond to
different values of the strangeness content of the proton,
$y = {2 \langle p|\bar{s}s|p \rangle / \langle p|\bar{u}u + \bar{d}d|p \rangle}
\approx 0$, 0.36\cite{dong}, 0.5\cite{gus},
in the chiral model.

We remark that the most recent lattice determination of the strangeness
content of the proton \cite{ohki}
(as well as recent theoretical results \cite{lyubovitskij})
indicate a very low value $y<0.05$,
in strong disagreement with previous calculations \cite{oldlat,dong,gus}.
If confirmed, such a small value would imply also a very small 
absolute value of $a_3$.
Using \cite{thorsson}
$ \langle p | \bar{d}d | p \rangle \approx \langle p | \bar{u}u | p \rangle
 = -(a_1+2a_3)$
and
$ \langle p | \bar{s}s | p \rangle = -2(a_2+a_3)$
we obtain
\be
 a_3 \approx { a_1y/2-a_2 \over 1-y } \gtrsim {-143\; \text{MeV} \over m_s} \;,
\ee
and kaon condensation would be strongly disfavored in the present model,
as will be illustrated below.

The thermodynamic potential densities due to the
condensed kaons and the thermal kaons are introduced as follows:
\bea
 \omega_{KN}^{c} &=& f_\pi^2 \left[
 \left({m_K^*}^2 -2b\mu_K\right)(1-\cos\theta)
 - \mu_K^2 {\sin^2\theta\over2} \right] \:,\quad
\\
 \omega_{KN}^\text{th} &=& T \intdp
 \ln\left[(1-e^{-\beta E^+_p})(1-e^{-\beta E^-_p})\right] \:.
\label{e:th}
\eea
Then the kaonic (charge) density $q_K$ is given by
\bea
 q_K &=& -{\partial\omega_{KN}\over\partial \mu_K}
\\&=&
 f_\pi^2 \Big[ 2b (1-\cos\theta) + \mu_K \sin^2 \theta  \Big]
\nonumber\\&&
 + \cos\theta \intdp \Bigl[ f_B(E^-_p) - f_B(E^+_p) \Bigr] \:,
\label{qk}
\eea
where the last term is the contribution due to thermally excited
kaons, $q_K^\text{th}$,
with the Bose distribution function $f_B(E)=1/(e^{\beta E}-1)$.

The kaon-nucleon free energy density
appearing in Eq.~(\ref{e:f})
obtained in this way is
\bea
 f_{KN} &=& \omega_{KN} + \mu_K q_K
\\ &=&
 f_\pi^2\left[
 {m_K^*}^2 (1-\cos\theta) + \mu_K^2 {\sin^2\theta\over2}
 \right]
\nonumber \\&&
 + \mu_K q_K^\text{th} + \omega_{KN}^\text{th} \:,
\label{ek}
\eea
and the internal energy density is
\bea
 \eps_{KN} &=& f_{KN} + T s_K,
\eea
where the kaonic entropy density is solely due to the thermal kaons:
\bea
 s_K &=& -{\partial\omega_{KN}\over\partial T}
 = \beta \left(  \eps_{KN}^\text{th} - \omega_{KN}^\text{th} \right)
\eea
with
\be
 \eps_{KN}^\text{th} =
 \intdp \Big[ E^-_p f_B(E^-_p) + E^+_p f_B(E^+_p) \Big] \:.
\ee

One can determine the ground state by minimizing the
total grand-canonical potential density $\omega_{KN}$
with respect to the condensate amplitude $\theta$, keeping $(\mu_K,\rho,x)$ fixed.
This minimization together with the chemical equilibrium and charge
neutrality conditions leads to the following three coupled
equations \cite{thorsson,kubis,tatsumi}
\bea
 0 &=& f_\pi^2\sin\theta
 \Big[ {m_K^*}^2  - 2b\mu_K - \mu_K^2\cos\theta \Big]
       + \frac{\partial\omega_{KN}^\text{th}}{\partial\theta} \:,
\label{e:th-min}
\eea
\bea
 \mu_K &=& \mu_n - \mu_p
\nonumber\\&=&
 4(1-2x){F_\text{sym}\over A} - (a_1m_s-\mu_K/2)(1-\cos\theta)
\nonumber\\&&
 - {1\over\rho}{\partial\omega_{KN}^\text{th} \over \partial x}
\:,
\label{e:betatheta}
\eea
\bea
 q_K+q_e+q_\mu = q_p = x\rho \:.
\label{e:cntheta}
\eea
Note that neglecting the thermal contribution in Eq.~(\ref{e:th-min})
implies $\widetilde \mu = \widetilde m$ and therefore $E_0^-=0$
in Eqs.~(\ref{e:epm}--\ref{e:muk}),
consistent with a singularity of the BE distribution function
and the existence of the condensate.
We therefore neglect also the thermal contribution in Eq.~(\ref{e:betatheta}),
as is done in Ref.~\cite{tatsumi}.

The lepton number density is given by ($l=e,\mu,\nu$):
\bea
 q_l = g_l \intdp
 \Big[ f_F(e_l({\bm p})-\mu_l) - f_F(e_l({\bm p})+\mu_l) \Big]
\label{e:ql}
\eea
with the Fermi distribution function $f_F(E)=1/(e^{\beta E}+1)$,
$e_l({\bm p})=\sqrt{m_l^2+p^2}$,
and the degeneracies $g_e=g_\mu=2$, $g_\nu=1$.

The composition and the EOS of the kaon-condensed phase in the chemically
equilibrated (P)NS matter can be obtained by solving the
coupled equations (\ref{e:th-min}), (\ref{e:betatheta}), and (\ref{e:cntheta}).
The critical density for kaon condensation is determined as the point
above which a real solution with $\theta>0$
for the coupled equations can be found.

\section{Composition and EOS of hot stellar matter}

In neutrino-trapped $\beta$-stable nuclear matter
the chemical potential of any particle $i=n,p,K,l$ is uniquely determined
by the conserved quantities baryon number $B_i$, electric charge $Q_i$,
and weak charges (lepton numbers) $L^{(e)}_i$, $L^{(\mu)}_i$:
\be
 \mu_i = B_i\mu_n - Q_i\mu_K
 + L^{(e)}_i\mu_{\nu_e}  + L^{(\mu)}_i\mu_{\nu_\mu} \:.
\label{mufre:eps}
\ee
For stellar matter containing nucleons, kaons, and leptons as relevant degrees
of freedom, the chemical equilibrium conditions read explicitly
\be
 \mu_K = \mu_n - \mu_p = \mu_e - \mu_{\nu_e} = \mu_\mu + \mu_{\bar{\nu}_\mu} \:.
\label{beta:eps}
\ee
At given baryon density $\rho$,
these equations have to be solved together with the
charge neutrality condition
\be
 \sum_i Q_i x_i = 0
\label{neutral:eps}
\ee
and those expressing conservation of lepton numbers
\be
 Y_l = x_l - x_{\bar l} + x_{\nu_l} - x_{\bar{\nu}_l}
 \:,\quad l=e,\mu \:.
\label{lepfrac:eps}
\ee

Gravitational collapse calculations of the electron-degenerate core of massive stars
indicate that at the onset of trapping, the electron lepton number is
$Y_e = x_e + x_{\nu_e} \approx 0.4$,
the precise value depending on the efficiency of electron capture
reactions during the initial collapse stage.
Also, because no muons are present when neutrinos become trapped,
the constraint
$Y_\mu = x_\mu - x_{\bar{\nu}_\mu} = 0$
is imposed.
We fix the $Y_l$ at these values in our calculations
for neutrino-trapped matter.
When the neutrinos have left the system,
their partial densities and chemical potentials vanish
and the above equations simplify accordingly.

The various chemical potentials are obtained from the total
free energy density $f$, Eq.~(\ref{e:f}),
\bea
 \mu_i(\{\rho_j\}) &=&
 \left. \frac{\partial f}{\partial \rho_i} \right|_{\rho_{j\neq i}} \:.
\label{mun:eps}
\eea
Once the hadronic and leptonic chemical potentials are known,
one can proceed to calculate the composition of the $\beta$-stable stellar matter,
and then the total pressure $p$ through the usual thermodynamical relation
\be
 p = \rho^2 {\partial{(f/\rho)}\over \partial{\rho}}
 = \sum_i \mu_i \rho_i - f  \:.
\ee

The stable configurations of a (P)NS can be obtained from the
well-known hydrostatic equilibrium equations
of Tolman, Oppenheimer, and Volkov \cite{shapiro}
for pressure $p(r)$ and enclosed mass $m(r)$
\bea
 {dp\over dr} &=& -\frac{Gm\eps}{r^2}
 \frac{\big( 1+ p/\eps \big) \big( 1 + 4\pi r^3p/m \big)}
 {1-2Gm/r} \:,
\label{tov1:eps}
\\
 \frac{dm}{dr} &=& 4\pi r^{2}\eps \:,
\label{tov2:eps}
\eea
once the EOS $p(\eps)$ is specified, with
$\eps=\eps_{NN}+\eps_{KN}+\eps_L$
the total internal energy density ($G$ is the gravitational constant).
For a chosen central value of the energy density, the numerical integration of
Eqs.~(\ref{tov1:eps}) and (\ref{tov2:eps}) provides the mass-radius relation.

Dynamical simulations of supernovae explosions \cite{burrows,pons,ponsevo}
show that the PNS has neither an isentropic nor an isothermal profile.
For simplicity we assume a constant temperature inside the star and
attach for the outer part a cold crust
given in Ref.~\cite{nv} for the medium-density regime
($0.001\;\mathrm{fm}^{-3}<\rho<0.08\;\mathrm{fm}^{-3}$),
and in Refs.~\cite{bps,fmt}
for the outer crust ($\rho<0.001\;\mathrm{fm}^{-3}$).
This schematizes the temperature profile of the PNS.
The other extreme choice of isentropic profiles has recently been investigated
within our approach \cite{isen}
and no significant qualitative differences have been found.

More realistic temperature profiles can be obtained by modeling
the neutrinosphere both in the interior and in the external outer
envelope, which is expected to be much cooler.
A proper treatment of the transition from the hot interior
to the cold outer part can
have a dramatic influence on the mass -- central density relation
in the region of low central density and low stellar masses.
In particular, the ``minimal mass" region,
typical of cold NS's \cite{shapiro},
can be shifted in PNS's to much higher values of central density and masses.
A detailed analysis of this point can be found in \cite{gondek},
where a model of the transition region
between the interior and the external envelope is developed.
However, the maximum mass region that we are interested in,
is hardly affected by the structure
of this low-density transition region \cite{nbbs}.

\section{Results}

\begin{figure}[t]
\includegraphics[height=110mm,angle=0]{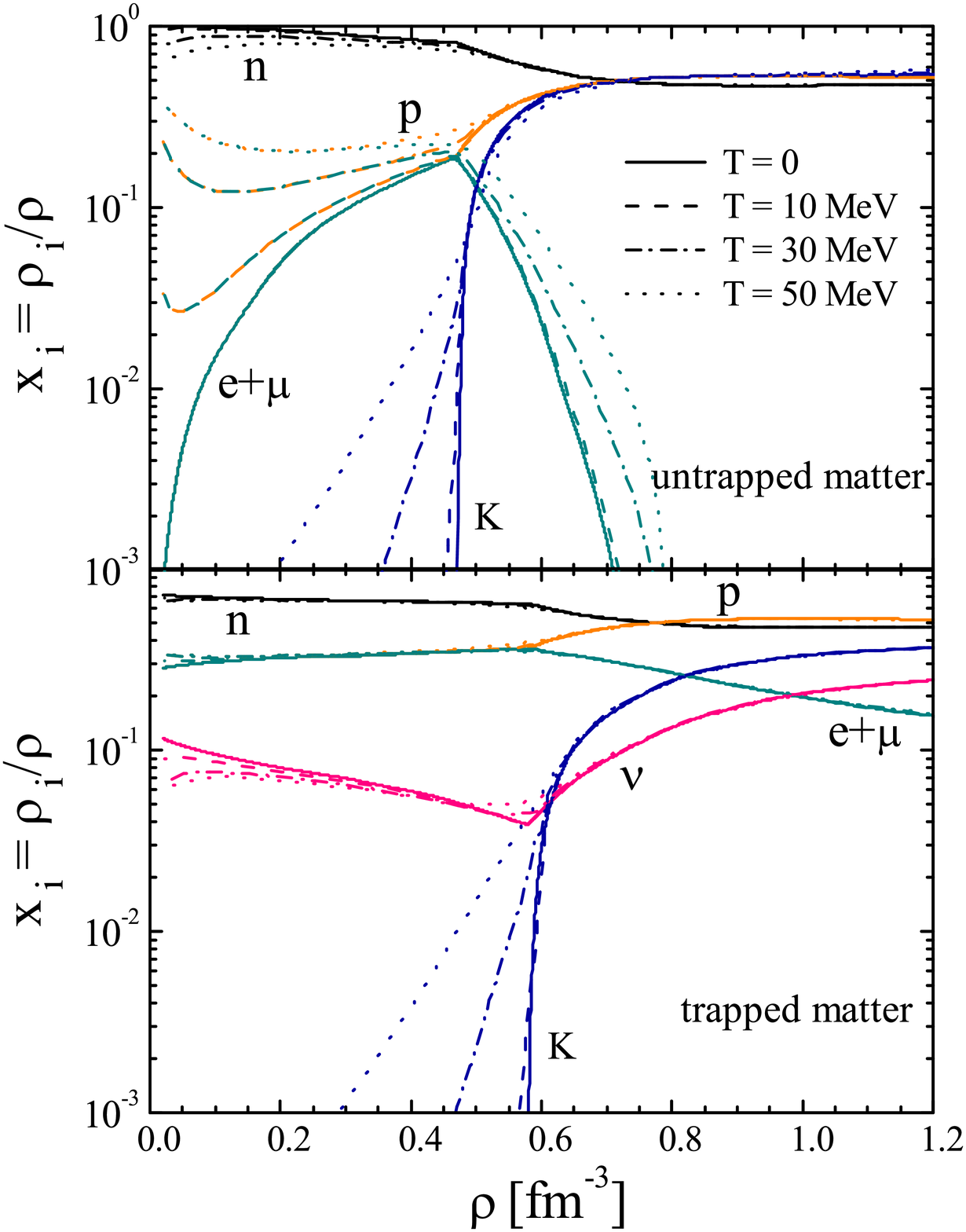}
\caption{(Color online)
Particle fractions as a function of the baryon density
in trapped ($Y_e=0.4$, lower panel)
and untrapped ($x_\nu=0$, upper panel) $\beta$-stable matter
at the temperatures $T$ = 0, 10, 30, and 50 MeV
for $a_3m_s = -222$ MeV and the micro TBF.}
\label{f:xt}
\end{figure}

\begin{figure*}[t]
\includegraphics[height=128mm,angle=0]{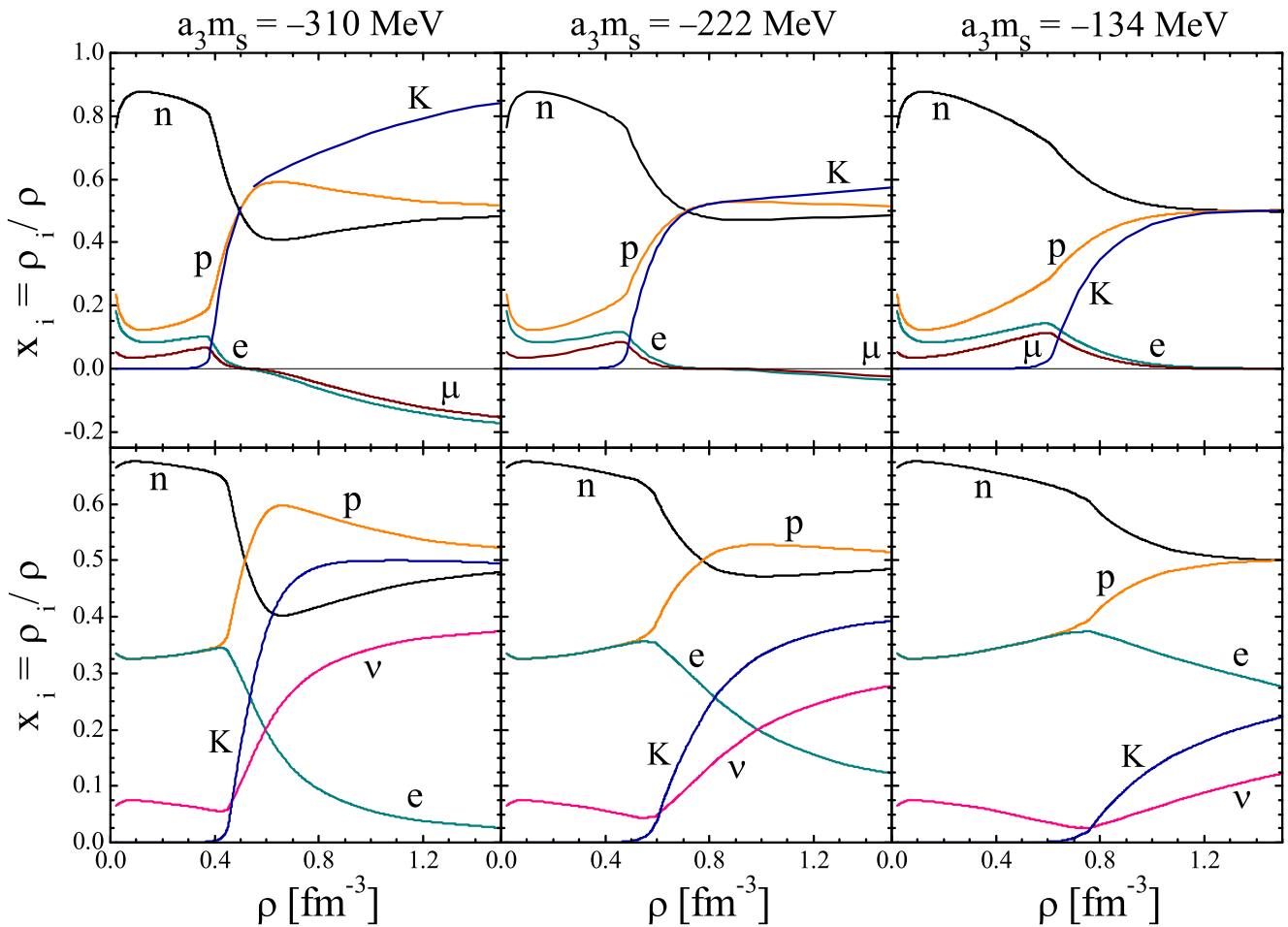}
\caption{(Color online)
Particle fractions as a function of the baryon density
in trapped ($Y_e=0.4$, lower panels)
and untrapped ($x_\nu=0$, upper panels) $\beta$-stable matter,
at temperature $T$ = 30 MeV and with the micro TBF
for $a_3m_s = -134,-222,-310$ MeV.
Negative values indicate an excess of antiparticles.
}
\label{f:xa}
\end{figure*}

\begin{figure*}[t]
\includegraphics[height=140mm,angle=0]{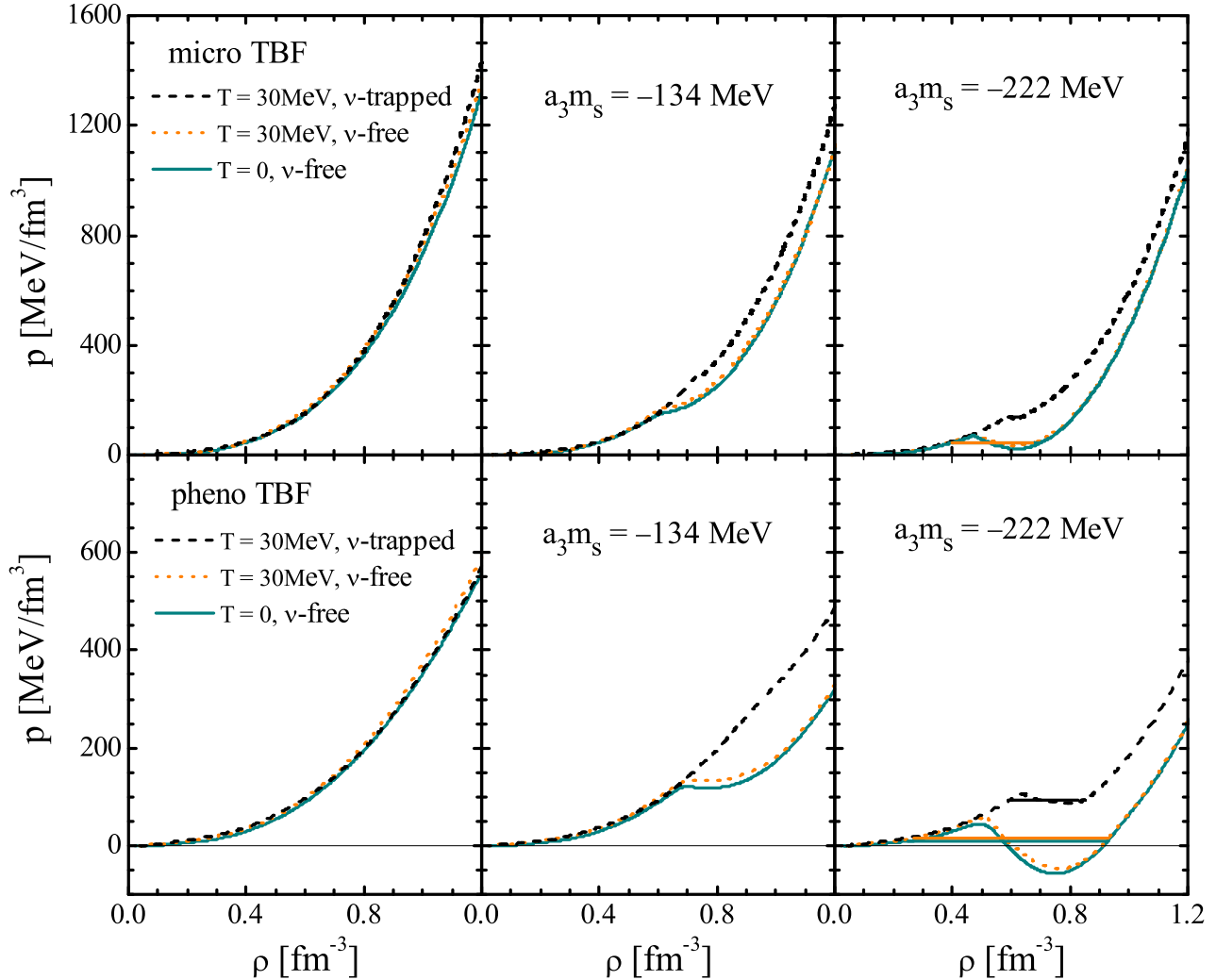}
\caption{(Color online)
The pressure of $\beta$-stable matter under the conditions
$T=30$ MeV, $Y_e=0.4$ (black dashed lines),
$T=30$ MeV, $x_\nu=0$ (red dotted lines),
$T=0$ MeV, $x_\nu=0$ (green solid lines),
is shown in the following cases:
no kaon condensate (left panels),
with kaon condensate and $a_3m_s = -134$ MeV (middle panels),
or $a_3m_s = -222$ MeV (left panels).
The upper (lower) panels display results obtained
with the micro (pheno) TBF.
Horizontal line segments illustrate the Maxwell constructions.}
\label{f:p}
\end{figure*}

\begin{figure*}[t]
\includegraphics[height=140mm,angle=0]{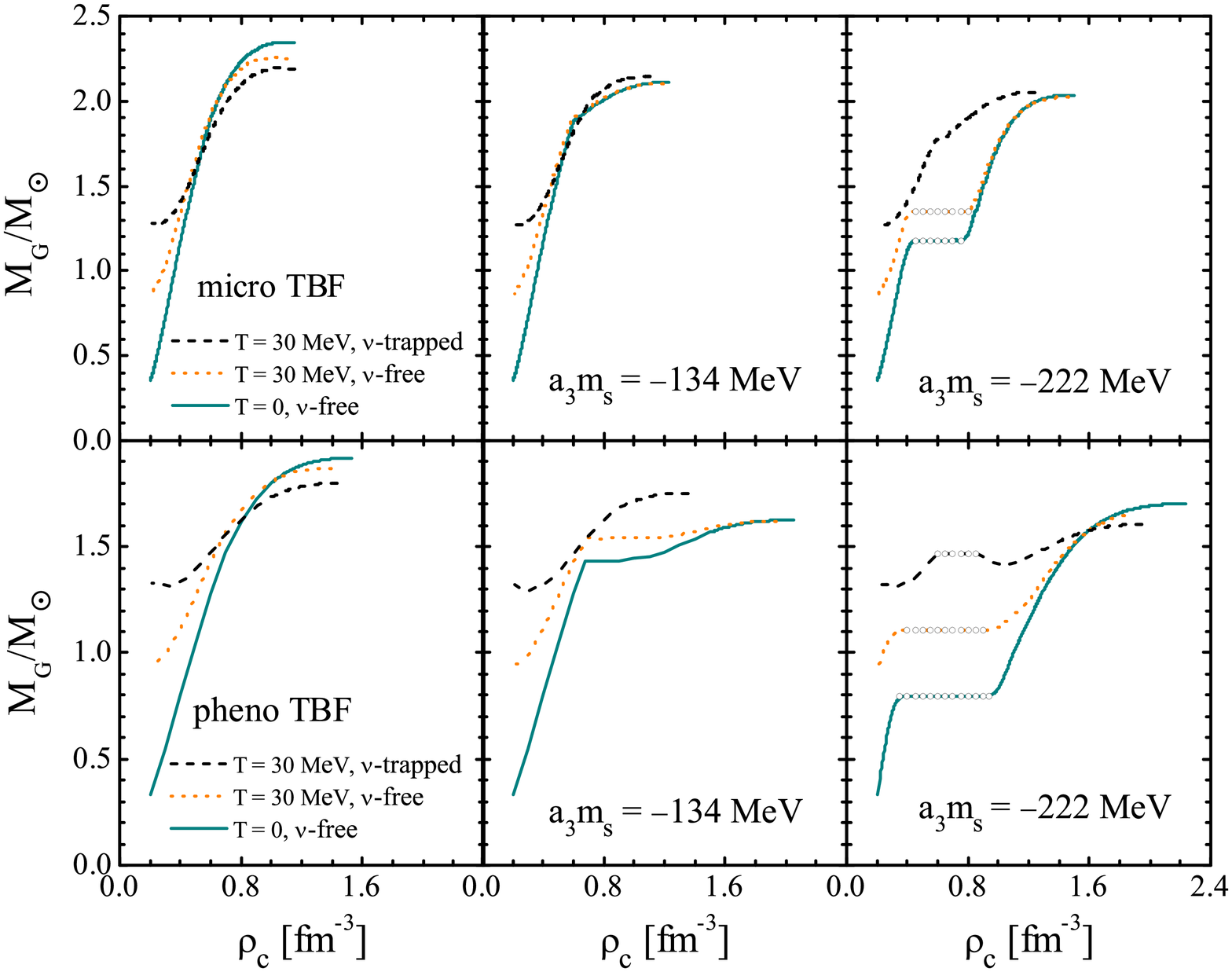}
\caption{(Color online)
(Proto)neutron star gravitational mass -- central density relations
obtained with the pheno TBF (upper panels)
and the micro TBF (lower panels)
under the conditions
$T=30$ MeV, $Y_e=0.4$ (black dashed lines),
$T=30$ MeV, $x_\nu=0$ (red dotted lines),
$T=0$ MeV, $x_\nu=0$ (green solid lines)
for $a_3m_s = -222$ MeV, $-134$ MeV, and without kaon condensation.}
\label{f:m}
\end{figure*}

In the following we present the results of our numerical calculations
regarding the composition of PNS matter and the structure of PNS's.

\subsection{Composition of stellar matter}

In Fig.~\ref{f:xt} we display the relative particle fractions
(of neutrons, protons, kaons, electrons, muons, and neutrinos)
in trapped (lower panel) and untrapped (upper panel) matter
as a function of the baryon density for several values of temperature
$T=0$, 10, 30, and 50 MeV,
obtained with $a_3m_s=-222$ MeV and the micro TBF.
We notice that temperature effects influence
the populations mainly in the low-density region,
and only slightly at high density.
Leptons are rather numerous at fairly small
densities as a result of Fermi distributions at finite temperature.

The kaon condensate threshold density is only slightly dependent
on the temperature,
namely 0.489, 0.490, 0.492, 0.497 fm$^{-3}$
at $T = 0$, 10, 30, 50 MeV for untrapped matter,
and 0.580, 0.583, 0.589, 0.629 fm$^{-3}$ for trapped matter, respectively.
The temperature influence on the kaon population is very small
above the condensate threshold and regards mainly the small fractions
of thermal kaons present before the threshold.
Above the critical density,
thermal effects increase the population of protons and leptons in the
untrapped case.
We remark that, as usually found \cite{thorsson,kubis}, in cold untrapped matter
the presence of a kaon condensate pushes the proton fraction above the
threshold allowing fast URCA cooling.

There is a large difference between untrapped and trapped matter,
where the kaon condensation sets in later, and the kaon concentration
remains lower.
The major reason is the smaller nuclear asymmetry of trapped matter,
which leads according to Eq.~(\ref{e:betatheta}) to a later kaon onset.
[The direct dependence of the kaon effective mass on the nuclear asymmetry,
Eq.~(\ref{e:mkstar}),
plays a minor role with the chosen interaction parameters].
In untrapped matter, the kaons replace immediately the leptons
in compensating the charge of the protons;
in trapped matter they cannot do that because the lepton number
has to be kept fixed.
Their effect is thus a moderate decrease of the charged leptons,
while the neutrino population increases.
Overall, their importance is substantially reduced compared
to the case of untrapped matter.

The dependence of the composition on the $KN$ interaction strength is
illustrated in Fig.~\ref{f:xa},
where we display the relative particle fractions
in trapped (lower panels) and untrapped matter (upper panels) at $T=30$ MeV
for the three different values of the interaction parameter $a_3$
that we consider.
The onset density of kaon condensation depends strongly on this parameter
and ranges approximately
from 0.4 fm$^{-3}$ to 0.6 fm$^{-3}$ in untrapped matter,
and from 0.45 fm$^{-3}$ to 0.75 fm$^{-3}$ in trapped matter.
The fairly large onset densities for $a_3m_s=-134$ MeV,
corresponding to a small strangeness content of the proton,
lie, however, in a region where the underlying concept of distinguishable
baryons and mesons becomes doubtful,
and also the simple chiral kaon-nucleon interaction would have to be
extended.

Regarding the dependence of the particle concentrations
on the TBF (micro or pheno) used,
it was shown in the zero-temperature calculations of Ref.~\cite{liang2}
that it is rather small,
with only some slight differences at high density,
where the micro TBF is stiffer than the pheno TBF.
We therefore do not repeat this comparison here,
but will only show the final (P)NS structure results
obtained with both TBF's in the next subsection.

\subsection{(Proto)neutron star structure}

Fig.~\ref{f:p} shows the EOS $p(\rho)$ obtained with the micro TBF (upper
panels) and the pheno TBF (lower panels) in the following cases:
(a) no kaon condensate (left panels),
(b) with kaon condensate using either $a_3m_s = -134$ MeV (middle panels)
or $a_3m_s = -222$ MeV (right panels).
We consider three different strongly idealized stages of the PNS evolution:
(i) $T=30$ MeV, $Y_e=0.4$ (black dashed lines),
the initial hot and neutrino-trapped state;
(ii) $T=30$ MeV, $x_\nu=0$ (red dotted lines),
the intermediate phase lasting about a few seconds,
when most neutrinos have diffused out of the still hot environment;
(iii) $T=0$ MeV, $x_\nu=0$ (green solid lines),
the final state of a cold NS formed after a few tens of seconds.
This rather crude treatment of the different stages of the PNS evolution
can obviously be improved once more realistic temperature/trapping profiles
become available. 
Fot the time being we consider it sufficient to reflect the gross 
qualitative features of the important evolution stages.

We observe that the kaon condensation produces a general softening of the
EOS with respect to the purely nucleonic case.
The degree of softening increases with the value of the
interaction parameter $|a_3m_s|$.
In the case with kaon condensate, neutrino trapping produces a stiffer
EOS due to the higher onset density of kaons and
smaller kaon abundance, as shown in Fig.~\ref{f:xa}.
This may lead a newly-formed, hot PNS to metastability,
i.e., a delayed collapse while cooling down, as discussed in
Refs.~\cite{rep,pons}.
One observes only a very small dependence of the EOS on the temperature,
which plays thus a minor role in comparison with neutrino trapping.
The above considerations hold true also when pheno TBF
are used in the baryonic EOS (lower panels of Fig.~\ref{f:p}),
where a softer increase of the pressure vs.~density
is observed as compared to the case with micro TBF.

In some cases, the onset of the kaon-condensed phase produces a
negative compressibility in the EOS.
Following Migdal \cite{migdal}, we have performed a Maxwell construction to maintain
a positive compressibility.
This implies the formation of a region of constant pressure,
comprised between two values of the baryon density,
whose extension depends on the magnitude of $|a_3m_s|$ \cite{kubis}.
It is indicated by horizontal lines in the relevant panels of Fig.~\ref{f:p}.

These features are reflected in Fig.~\ref{f:m},
where the corresponding gravitational mass -- central density relations are plotted.
The upper (lower) panels show results obtained with micro (pheno) TBF,
following the same notation as in Fig.~\ref{f:p}.
In the case without kaons (left panels),
the maximum mass of the PNS is slightly smaller than that of the NS,
because neutrino trapping reduces the asymmetry of beta-stable matter.
The presence of kaon condensation reverses the situation,
and the PNS generally has a larger maximum mass than the NS,
due to the less softening effect of kaons in trapped matter.
A delayed collapse scenario is therefore facilitated by the presence of
a kaon condensate, as is indeed generally found \cite{thorsson,tatsumi,pons}.
We remind that a similar effect is also produced when hyperons are introduced
in beta-stable stellar matter \cite{nbbs}.

Again the effect of finite temperature is minor compared to the one of trapping, 
so that a heavy PNS would be destabilized 
by loss of neutrinos early during its evolution.
Regarding this feature,
similar conclusions as ours can be drawn from the results of 
Refs.~\cite{rep,tatsumi}, whereas
somewhat larger effects of finite temperature have been claimed in Ref.~\cite{pons}.
However, in that case also a higher typical temperature was assumed
in stage (ii) than in stage (i) of our evolution scenario, 
which accounts at least for part of the observed differences.
It is clearly desireable to use a more realistic temperature profile
for a more reliable evaluation of this feature in the future.

A rather extreme scenario is seen in the case of a quite soft nuclear EOS
combined with a strong kaon condensate
(lower right panel of Fig.~\ref{f:m}),
where actually the maximum mass of the NS remains higher than that of the PNS
and no delayed collapse could occur.
A further consequence is the occurrence of gravitationally unstable sections
in the mass -- central density plot (see the black dashed curve),
which have also been observed in Refs.~\cite{thorsson,tatsumi}.
However, we consider this combination of extreme parameter choices unlikely,
as it requires a very large density jump in the Maxwell construction
(thus obliterating the soft part of the EOS)
and also leads to unrealistically high central densities of the star.
Furthermore, the case of a strong kaon condensate seems to be excluded
in the present model now, as discussed before.

The global properties of the different configurations of (P)NS are summarized
in Table~\ref{t:mass}.
One notes that the theoretical predictions for the maximum masses depend most 
importantly on the nuclear EOS, whereas 
the effects of kaon condensation and/or neutrino trapping are of smaller magnitude.
Current observational values \cite{nsobs} suggest the existence of stars
heavier than about 1.7 $M_\odot$, although their accurate confirmation
is still eagerly awaited. 
Even if they are, somewhat higher values of about 2 $M_\odot$ 
would be required in order to really discriminate between different nuclear EOS.

\begin{table}[b]
\caption{Properties of (proto)neutron stars.}
\medskip
\begin{ruledtabular}
\begin{tabular}{l|c|dd|dd}
 & & \multicolumn{2}{c|}{micro TBF} & \multicolumn{2}{c}{pheno TBF} \\
 & $a_3m_s$ (MeV)
 & \multicolumn{1}{c}{$M_\text{max}/M_\odot$} & \multicolumn{1}{c|}{$\rho_c/\rho_0$}
 & \multicolumn{1}{c}{$M_\text{max}/M_\odot$} & \multicolumn{1}{c}{$\rho_c/\rho_0$} \\
\hline
 trapped     & ----  & 2.19 & 6.29 & 1.80 & 8.24\\
 $T=30$ MeV    & -134  & 2.14 & 6.24 & 1.75 & 7.82\\
             & -222  & 2.05 & 7.12 & 1.61 & 11.47\\
\hline
 untrapped   & ----  & 2.26 & 6.00 & 1.87 & 7.94\\
 $T=30$ MeV    & -134  & 2.10 & 6.76 & 1.62 & 11.17\\
             & -222  & 2.02 & 8.35 & 1.67 & 12.53\\
\hline
 untrapped   & ----  & 2.34 & 6.29 & 1.92 & 8.76\\
 $T=0$         & -134  & 2.11 & 6.88 & 1.62 & 11.76\\
             & -222  & 2.03 & 8.47 & 1.70 & 12.94\\
\end{tabular}

\end{ruledtabular}
\label{t:mass}
\end{table}

\section{Summary}

In conclusion, we presented microscopic calculations   
and convenient parametrizations of the equation of state 
of hot asymmetric nuclear matter within the framework of the
Brueckner-Hartree-Fock approach with two different nuclear three-body forces.
We then investigated the EOS as well as the consequences
of including kaon condensation in hot and neutrino-trapped NS matter,
employing a standard chiral model at finite temperature.
Effects of finite temperature are thus included consistently in both the 
nucleonic and the kaonic part of the interaction.

Our results are qualitatively in agreement with those obtained with more 
phenomenological approaches \cite{rep,tatsumi},
although the quantitative predictions turn out to be different. 
In particular we found that also in our microscopic approach 
finite temperature plays a minor role compared to neutrino trapping,
which generally decreases the stellar maximum mass
in the absence of a kaon condensate, and increases it with a condensate.
This is due to the reduced appearance of kaons in trapped vs.~untrapped matter.
Global PNS properties are, however, determined primarily by the {\em nucleonic} 
part of the EOS.

Furthermore, if recent very small values for the strangeness content of
the proton are confirmed, kaon condensation in the present model
sets in only at a critical density
$\rho^K_c \gtrsim 4\rho_0$,
whereas in the same BHF framework hyperons appear at
$\rho^Y_c \approx (2-3)\rho_0$ \cite{litbf,bbs,hypns},
and would then completely suppress the kaon degree of freedom.
In any case, the maximum mass of a (P)NS is strongly reduced by the
appearance of strangeness in the relevant dense environment,
either in the form of kaons, or of hyperons;
and, in both cases a delayed collapse scenario appears very probable.

\section*{Acknowledgments}

This work is supported in part by the National Natural Science
Foundation of China (10605018,10905048,10975116),
the Knowledge Innovation Project (KJCX3-SYW-N2) of the Chinese Academy of Sciences, 
the Program for New Century Excellent Talents in University (NCET-07-0730), 
the Asia-Europe Link project (CN/ASIA-LINK/008(94791)) of the European Commission,
and by COMPSTAR, a research networking program of the European Science Foundation.



\begin{thebibliography}{99}

\bibitem{kaplan}
 D. B. Kaplan and A. E. Nelson,
 Phys. Lett. {\bf B175}, 57 (1986);
 Nucl. Phys. {\bf A479}, 273 (1988).

\bibitem{politzer}
 H. D. Politzer and M. B. Wise,
 Phys. Lett. {\bf B273}, 156 (1991).

\bibitem{brownevo}
 G. E. Brown, K. Kubodera, M. Rho, and V. Thorsson,
 Phys. Lett. {\bf B291}, 355 (1992).
\bibitem{brown}
 C. H. Lee, G. E. Brown, D. P. Min, and M. Rho,
 Nucl. Phys. {\bf A585}, 401 (1995);
 C. H. Lee,
 Phys. Rep. {\bf 275}, 255 (1996)
 and references therein;
 G. E. Brown, C. H. Lee, and M. Rho,
 Phys. Rep. {\bf 462}, 1 (2008).

\bibitem{thorsson}
 V. Thorsson, M. Prakash, and J. M. Lattimer,
 Nucl. Phys. {\bf A572}, 693 (1994);
 {\bf A574}, 851 (1994), Erratum.

\bibitem{ellis}
 P. J. Ellis, R. Knorren, and M. Prakash,
 Phys. Lett. {\bf B349}, 11 (1995).

\bibitem{glendenning}
 N. K. Glendenning and J. Schaffner-Bielich,
 Phys. Rev. Lett. {\bf 81}, 4564 (1998);
 Phys. Rev. {\bf C60}, 025803 (1999).

\bibitem{rep}
 M. Prakash, I. Bombaci, M. Prakash, P. J. Ellis, J. M. Lattimer, and R. Knorren,
 Phys. Rep. {\bf 280}, 1 (1997).

\bibitem{tatsumi}
 T. Tatsumi and M. Yasuhira,
 Phys. Lett. {\bf B441}, 9 (1998);
 Nucl. Phys. {\bf A653}, 133 (1999);
 M. Yasuhira and T. Tatsumi,
 Nucl. Phys. {\bf A690}, 769 (2001);
 T. Muto, M. Yasuhira, T. Tatsumi, and N. Iwamoto,
 Phys. Rev. {\bf D67}, 103002 (2003).

\bibitem{tatsumievo}
 T. Muto, T. Tatsumi, and N. Iwamoto,
 Phys. Rev. {\bf D61}, 063001,083002 (2000).

\bibitem{pons}
 J. A. Pons, S. Reddy, P. J. Ellis, M. Prakash, and J. M. Lattimer,
 Phys. Rev. {\bf C62}, 035803 (2000);
 J. A. Pons, J. A. Miralles, M. Prakash, and J. M. Lattimer,
 Astrophys. J. {\bf 553}, 382 (2001).

\bibitem{ramos}
 A. Ramos, J. S. Bielich, and J. Wambach,
 Lect. Notes. Phys. {\bf 578}, 175 (2001).

\bibitem{carlson}
 J. Carlson, H. Heiselberg, and V. R. Pandharipande,
 Phys. Rev. {\bf C63}, 017603 (2000).

\bibitem{norsen}
 T. Norsen and S. Reddy,
 Phys. Rev. {\bf C63}, 065804 (2001).

\bibitem{kubis}
 S. Kubis and M. Kutschera,
 Nucl. Phys. {\bf A720}, 189 (2003).

\bibitem{liang1}
 W. Zuo, A. Li, Z. H. Li, and U. Lombardo,
 Phys. Rev. {\bf C70}, 055802 (2004).
\bibitem{liang2}
 A. Li, G. F. Burgio, U. Lombardo, and W. Zuo,
 Phys. Rev. {\bf C74}, 055801 (2006).

\bibitem{oldlat}
 M. Fukugita, Y. Kuramashi, M. Okawa, and A. Ukawa,
 Phys. Rev. {\bf D51}, 5319 (1995).

\bibitem{dong}
 S. J. Dong, J.-F. Laga\"e, and K. F. Liu,
 Phys. Rev. {\bf D54}, 5496 (1996).

\bibitem{gus}
 S. G\"usken et al.,
 Phys. Rev. {\bf D59}, 054504 (1999).

\bibitem{lyubovitskij}
 V. E. Lyubovitskij, T. Gutsche, A. Faessler, and E. G. Drukarev,
 Phys. Rev. {\bf D63}, 054026 (2001).

\bibitem{ohki}
 H. Ohki et al.,
 Phys. Rev. {\bf D78}, 054502 (2008).

\bibitem{schaffner}
 R. Knorren, M. Prakash, and P. J. Ellis,
 Phys. Rev. {\bf C52}, 3470 (1995);
 J. Schaffner-Bielich and I. N. Mishustin,
 Phys. Rev. {\bf C53}, 1416 (1996);
 T. Muto,
 Phys. Rev. {\bf C77}, 015810 (2008).

\bibitem{baum}
 T. W. Baumgarte, S. L. Shapiro, and S. A. Teukolsky,
 Ap. J. {\bf 458}, 680 (1996).

\bibitem{taka}
 T. Takatsuka,
 Prog. Theor. Phys. {\bf 95}, 901 (1996).

\bibitem{schaab}
 K. Strobel, C. Schaab, M. K. Weigel,
 Astron. Astrophys., {\bf 350}, 497 (1999);
 K. Strobel, M. K. Weigel,
 Astron. Astrophys., {\bf 367}, 582 (2001).

\bibitem{nbbs}
 O. E. Nicotra, M. Baldo, G. F. Burgio, and H.-J. Schulze,
 Astron. Astrophys. {\bf 451}, 213 (2006);
 O. E. Nicotra, M. Baldo, G. F. Burgio, and H.-J. Schulze,
 Phys. Rev. {\bf D74}, 123001 (2006).

\bibitem{isen}
 G. F. Burgio and H.-J. Schulze,
 Physics of Atomic Nuclei {\bf 72}, 1197 (2009).

\bibitem{lej}
 A. Lejeune, P. Grang\a'e, M. Martzolff, and J. Cugnon,
 Nucl. Phys. {\bf A453}, 189 (1986).

\bibitem{bombaci}
 I. Bombaci and U. Lombardo,
 Phys. Rev. {\bf C44}, 1892 (1991);
 W. Zuo, I. Bombaci, and U. Lombardo,
 Phys. Rev. {\bf C60}, 024605 (1999).

\bibitem{book}
 M. Baldo,
 {\em Nuclear Methods and the Nuclear Equation of State},
 International Review of Nuclear Physics, Vol. 8
 (World Scientific, Singapore, 1999).

\bibitem{baldo}
 M. Baldo and L. S. Ferreira,
 Phys. Rev. {\bf C59}, 682 (1999).

\bibitem{wiringa}
 R. B. Wiringa, V. G. J. Stoks, and R. Schiavilla,
 Phys. Rev. {\bf C51}, 38 (1995).

\bibitem{grange}
 P. Grang\'e, A. Lejeune, M. Martzolff, and J.-F. Mathiot,
 Phys. Rev. {\bf C40}, 1040 (1989).

\bibitem{zuotbf}
 W. Zuo, A. Lejeune, U. Lombardo, and J.-F. Mathiot,
 Nucl. Phys. {\bf A706}, 418 (2002);
 Z. H. Li, U. Lombardo, H.-J. Schulze, and W. Zuo,
 Phys. Rev. {\bf C77}, 034316 (2008).

\bibitem{litbf}
 Z. H. Li and H.-J. Schulze,
 Phys. Rev. {\bf C78}, 028801 (2008).

\bibitem{uix}
 J. Carlson, V. R. Pandharipande, and R. B. Wiringa,
 Nucl. Phys. {\bf A401}, 59 (1983);
 R. Schiavilla, V. R. Pandharipande, and R. B. Wiringa,
 Nucl. Phys. {\bf A449}, 219 (1986).

\bibitem{bbb}
 M. Baldo, I. Bombaci, and G. F. Burgio,
 Astron. Astrophys. {\bf 328}, 274 (1997);
 X. R. Zhou, G. F. Burgio, U. Lombardo, H.-J. Schulze, and W. Zuo,
 Phys. Rev. {\bf C69}, 018801 (2004).

\bibitem{bbs}
 M. Baldo, G. F. Burgio, and H.-J. Schulze,
 Phys. Rev. {\bf C58}, 3688 (1998).

\bibitem{shapiro}
 S. L. Shapiro and S. A. Teukolsky,
 {\em Black Holes, White Dwarfs, and Neutron Stars}
 (John Wiley and Sons, New York, 1983).

\bibitem{burrows}
 A. Burrows and J. M. Lattimer,
 Astrophys. J. {\bf 307}, 178 (1986).

\bibitem{ponsevo}
 J. A. Pons, S. Reddy, M. Prakash, J. M. Lattimer, and J. A. Miralles,
 Astrophys. J. {\bf 513}, 780 (1999);
 L. Villain, J. A. Pons, P. Cerd\'a-Dur\'an, and E. Gourgoulhon,
 Astron. Astrophys. {\bf 418}, 283 (2004).

\bibitem{nv}
 J. W. Negele and D. Vautherin,
 Nucl. Phys. {\bf A207}, 298 (1973).
\bibitem{bps}
 G. Baym, C. Pethick, and D. Sutherland,
 Astrophys. J. {\bf 170}, 299 (1971).
\bibitem{fmt}
 R. Feynman, F. Metropolis, and E. Teller,
 Phys. Rev. {\bf C75}, 1561 (1949).

\bibitem{gondek}
 D. Gondek, P. Haensel, and J. L. Zdunik,
 Astron. Astrophys. {\bf 325}, 217 (1997).

\bibitem{migdal}
 A. B. Migdal,
 in {\em Mesons in Nuclei}, vol.~3, eds. M. Rho and D. Wilkinson
 (North-Holland, Amsterdam, 1979).

\bibitem{nsobs}
 J. M. Lattimer and M. Prakash,
 Phys. Rep. {\bf 442}, 109 (2007).

\bibitem{hypns}
 M. Baldo, G. F. Burgio, and H.-J. Schulze,
 Phys. Rev. {\bf C61}, 055801 (2000);
 H.-J. Schulze, A. Polls, A. Ramos, and I. Vida\a~na,
 Phys. Rev. {\bf C73}, 058801 (2006).

\end{thebibliography}
\end{document}